\documentclass[conference]{IEEEtran}

\usepackage[cmex10]{amsmath}
\usepackage[cmex10]{amsmath}
\usepackage{array}
\usepackage{verbatim}

\usepackage{tikz}
\usepackage{epstopdf}
\usepackage{amsfonts}
\usepackage{amsmath}
\usepackage{amssymb}
\newtheorem{defi}{Definition}
\newtheorem{theorem}{Theorem}
\newtheorem{lemma}{Lemma}

\newtheorem{proposition}{Proposition}

\newcommand{\sm}{\sum _{\ell=1}^K}

\newcommand{\modnl}{\hspace{-0.1in}\mod\Lambda_{\ell}}

\newcommand{\modla}{\hspace{-0.1in}\mod\Lambda_{1}}

\newcommand{\rf}{\left(\frac{r_{\mathrm{eff},\ell}}{r_{\mathrm{cov},\ell}}\right)}
\usepackage{cite}
\usepackage[center]{caption}
\usepackage{graphicx}
\usepackage{multicol}
\begin{document}
\title{Compute-and-Forward Can Buy Secrecy Cheap}

\author{\IEEEauthorblockN{Parisa Babaheidarian*, Somayeh Salimi**}
\IEEEauthorblockA{
*Boston University,**KTH Royal Institute of Technology}}

\maketitle
\begin{abstract}
We consider a Gaussian multiple access channel with $K$~transmitters, a (intended) receiver and an external eavesdropper. The transmitters wish to reliably communicate with the receiver while concealing their messages from the eavesdropper. This scenario has been investigated in prior works using two different coding techniques; the random i.i.d. Gaussian coding and the signal alignment coding. Although, the latter offers promising results in a very high SNR regime, extending these results to the finite SNR regime is a challenging task. In this paper, we propose a new lattice alignment scheme based on the compute-and-forward framework which works at any finite SNR. We show that our achievable secure sum rate scales with $\log(\mathrm{SNR})$ and hence, in most SNR regimes, our scheme outperforms the random coding scheme in which the secure sum rate does not grow with power. Furthermore, we show that our result matches the prior work in the infinite SNR regime. Additionally, we analyze our result numerically.
\end{abstract}
\vspace{-2.4mm}
\section{Introduction}
\vspace{-0.9mm}
Gaussian Multiple Access Channel (MAC) has been considered under different security scenarios. One interesting scenario is the $K$-user Gaussian MAC with an external eavesdropper in which the users wish to reliably send their messages to the receiver while keeping them hidden from the eavesdropper. This scenario has been investigated in~\cite{tekin2008general} using the Gaussian i.i.d. random codes. Although, these codes achieve the capacity region of MAC without security, the result in~\cite{tekin2008general} shows that they have a poor performance in relatively high SNR regimes when the security constraint is added. In an attempt to improve the high SNR results, researchers investigated the problem using the signal alignment technique. In particular, in~\cite{bagherikaram2 010secure} and\cite{xie2012secure}, it is shown that their proposed schemes offer a significant improvement over the random coding counterpart in a very high SNR regime. In fact, the scheme proposed in~\cite{xie2012secure} achieves the optimal secure Degrees of Freedom (DoF) of the $K$-user Gaussian wiretap MAC. However, as these alignment schemes use a maximum-likelihood decoder, bounding the error probability of the decoder in the finite SNR regime is challenging and this limits their results to the high SNR regime.\\
\indent In light of lattice alignment technique, the compute-and-forward framework was proposed in~\cite{nazer2011compute} which can operate at any finite SNR. Recently, the $K$-user Gaussian MAC without security constraint has been investigated in~\cite{ordentlich2012approximate} based on lattice coding and the compute-and-forward framework. The proposed scheme in\cite{ordentlich2012approximate} achieves the MAC sum capacity within a constant gap and for any finite SNR.\\
\indent Motivated by the above arguments, we propose a new achievability scheme for the $K$-user Gaussian wiretap MAC in which lattice alignment is used along with the asymmetric compute-and-forward framework. We evaluate the performance of our proposed scheme both analytically and numerically for any finite SNR. We prove that our proposed scheme achieves a secure sum rate that scales with $\log(\mathrm{SNR})$, in contrast to the Gaussian random coding result which does not grow with SNR and therefore, it somehow fails at moderate and high SNR regimes. Finally, we show that the asymptotic behavior of our proposed scheme agrees with the prior work result in~\cite{bagherikaram2010secure} in the high SNR regime.\\
\indent The paper is organized as follows. In Section II, our setup preliminaries are described. Our main result is given in Section
III along with the comparison to the prior works. In Section IV, the proof of the main result is presented. We conclude the paper in Section V. The proof of Lemma 1 used in Section IV is given in Appendix.
\vspace{-2.5mm}
\section{Problem Statement}\label{sec2}
\vspace{-0.3mm}
A $K$-user (real) Gaussian wiretap multiple access channel (MAC) consists of $K$ transmitters, a receiver and an external eavesdropper. The relations between the channel inputs and outputs are given as
\vspace{-2.6mm}
\begin{equation} \label{c1}
\mathbf{y}=\sm h_{\ell}\mathbf{x}_{\ell}+\mathbf{z}, \quad \mathbf{y}_E=\sm g_{\ell}\mathbf{x}_{\ell}+\mathbf{z}_E
\end{equation}
\vspace{-0.5mm}
where $\mathbf{x}_{\ell}$ is an $N$-length channel input vector of user~$\ell$ which satisfies the following power constraint.
\vspace{-2mm}
\begin{equation}
\|\mathbf{x_{\ell}}\|^2 \leq NP \label{power},~~ \forall \ell \in\{1,\dots,K\}
\end{equation}
The vectors $\mathbf{y}$ and $\mathbf{y}_E$ in (\ref{c1}) are the receiver and the eavesdropper channel outputs, respectively. Also, $\mathbf{z}$ and $\mathbf{z}_E$ are the independent channel noises, each distributed i.i.d. according to  $\mathcal{N}(0,1)$. Finally, vectors $\mathbf{h}\triangleq[h_1,\dots,h_K]^{T}$ and $\mathbf{g}\triangleq[g_1,\dots,g_K]^{T}$ are real-valued vectors representing the channel gains to the receiver and the eavesdropper, respectively. The channel model is illustrated in Fig. 1.\\
\indent User~$\ell$ encodes its confidential message $W_{\ell}$, which is uniformly distributed over the set~$\{1,\dots, 2^{NR_{\ell}}\}$ and is independent of other users' messages, through some stochastic mapping~$\mathcal{E}_{\ell}$, i.e., $\mathbf{x}_{\ell} = \mathcal{E}_{\ell}(W_{\ell})$, for $\ell \in \{1,\dots,K\}$. There is also a decoder~$D$ at the receiver side which estimates the messages, i.e., $D(\mathbf{y})=\{\hat{W}_{\ell}\}_{\ell=1}^{K}$.
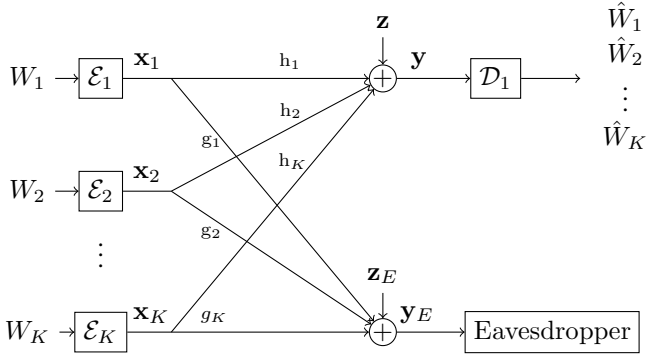
\begin{figure}
\centering
\begin{tikzpicture}[scale=0.75]
\draw(0.7,6) node (nodew1) {$W_1$};
\draw (2,6) node[draw] (nodeE1) {$\mathcal E_1$};
\draw (7,6) node[circle,draw,inner sep=0] (nodeplus1) {$+$};
\draw (7,7) node (nodenoise1) {$\mathbf z$};
\draw (9,6) node[draw] (nodeD1) {$\mathcal D_1$};

\draw(11.3,6) node (nodewhat1) {$\begin{array}{c}
 \hat W_1 \\ \hat W_2 \\ \vdots \\ \hat W_K \end{array} $
 };
\draw[->] (nodew1)--(nodeE1);
\draw[->] (nodeE1)--(nodeplus1) node[pos=0.1,sloped,above]{$\mathbf{x}_1$} node[pos=0.68,sloped,above] {$\scriptstyle \mathrm h_1$};
\draw[->] (nodenoise1)--(nodeplus1);
\draw[->] (nodeplus1)--(nodeD1) node[pos=0.3,sloped,above]{$\mathbf y$};
\draw[->] (nodeD1)--(nodewhat1);
\draw(0.7,4) node (nodew2) {$W_2$};
\draw (2,4) node[draw] (nodeE2) {$\mathcal E_2$};
\draw[->] (nodew2)--(nodeE2);
\draw (nodeE2)--(3.25,4) node[pos=0.5,sloped, above]{$\mathbf{x}_2$};
\draw(0.7,1.5) node (nodew3) {$W_K$};
\draw (2,1.5) node[draw, minimum size = 10] (nodeE3) {$\mathcal E_K$};
\draw (7,1.5) node[circle,draw,inner sep=0] (nodeplus3) {$+$};
\draw (7,2.5) node (nodenoise3) {$ \mathbf{z}_E$};
\draw (10,1.5) node[draw] (nodeD3) {$ \mathrm {Eavesdropper}$};
\draw[->] (nodew3)--(nodeE3);
\draw[->] (nodeE3)--(nodeplus3)node[pos=0.1,sloped,above]{$\mathbf{x}_K$} node[pos=0.36,sloped,above] {$\scriptstyle g_K$};
\draw[->] (nodenoise3)--(nodeplus3);
\draw[->] (nodeplus3)--(nodeD3) node[pos=0.3,sloped,above]{$\mathbf y_E$};
\draw (2,3) node {$\vdots$};
\draw[->] (3.25,6)--(nodeplus3) node[pos=0.2,below]{$\scriptstyle \mathrm g_1$};

\draw[->] (3.25,4)--(nodeplus1) node[pos=0.6,above]{$\scriptstyle \mathrm h_2$};
\draw[->] (3.25,4)--(nodeplus3) node[pos=0.2,below]{$\scriptstyle \mathrm g_2$};
\draw[->] (3.25,1.5)--(nodeplus1) node[pos=0.6,above, yshift=3]{$\scriptstyle \mathrm h_K$};
\end{tikzpicture}
\vspace{0.02 in}
\caption{ \small {The asymmetric Gaussian wiretap multiple access channel model.}}
\vspace{-5mm}
\end{figure}
\begin{defi}[Achievable secure sum rate]\label{achievability tuple}
For~the described channel model, a secure sum rate~$\sm R_{\ell}$ is achievable, if for any$~\epsilon>0$ and large enough $N$, there exist a sequence of encoders $\{\mathcal{E}_{\ell}\}_{\ell=1}^{k}$ and a decoder $D$ such that
\vspace{-2.6mm}
\begin{equation}
\mathrm{Pr}\left(\bigcup_{\ell=1}^K\lbrace \hat{W}_{\ell} \neq W_{\ell} \rbrace\right)< \epsilon \label{c2}
\end{equation}
\vspace{-2.5mm}
\begin{equation}
\sm R_{\ell}\leq \frac{1}{N}H(W_1,W_2,\dots,W_K|\mathbf{y}_E)+\epsilon \label{c3}
\end{equation}
where $\mathrm{Pr}$ denotes the probability of the event.\footnote{Note that in Definition~1 we are interested in weak secrecy.} The secure sum capacity is the supremum of all achievable secure sum rates.
\end{defi}
\vspace{-1.5mm}
\section{Main Results}\label{sec4}
The problem described in Section~\ref{sec2} has been treated in \cite{bagherikaram2010secure} and \cite{xie2012secure} in the infinite SNR regime. Their proposed schemes is based on bounding the minimum distance between the codewords in the receiver's effective codebook. Using this method, they showed that the decoding error probability tends to zero, provided that the input power goes to infinity. In this paper, we present a new scheme which provides a lower bound on the secure sum capacity for the same model and for any finite value of SNR. To this end, we utilize the compute-and-forward framework presented in~\cite{nazer2011compute}. More precisely, we develop a coding scheme using an asymmetric compute-and-forward framework to address the asymmetric transmitter-eavesdropper channel gains, i.e., different values of $g_\ell$ for different users. It should be noted that the asymmetric compute-and-forward framework is also treated in~\cite{ntranos2013usc}, but here we add the security constraint to the framework.\\
\indent In the compute-and-forward framework, the receiver first decodes $K$ linearly independent integer combinations of the transmitted lattice codewords and then, it solves the equations for its desired lattice codewords.\footnote{The rates are determined by how closely the equations integer coefficients match the channel gains $\mathbf{h}_{\ell}$.} The equations are decoded successively meaning that at each step $k$, the receiver cancels the effect of the $k-1$ previously decoded codewords from the current equation and solves it for the next codeword. The approach is similar to the Gaussian elimination with a difference that row switching is not allowed here. This limitation is due to the fact that a codeword cannot be eliminated from the current equation using another equation which has not been decoded yet. As a result, the order of canceling out the codewords cannot be chosen arbitrarily, however, it can be shown that there exists at least one successive cancellation order such that all~$K$ codewords can be decoded~\cite{ordentlich2012approximate}.
\begin{proposition}
Consider an index permutation function $\pi$, i.e., $\pi:\{1,\dots,K\}\rightarrow \{1,\dots,K\}$, which gives a successive cancellation order in the compute-and-forward framework. Also, assume that the set of linearly independent integer-valued $K$-length vectors~$\{\mathbf{a}_1,\dots,\mathbf{a}_K\}$ be the equations coefficients. Then, for the channel model in Section~\ref{sec2}, the receiver can decode the message $W_{\ell} \in \{1,\dots,2^{NR_{\ell}}\}$ with a vanishing error probability if
\vspace{-2mm}
\begin{equation}
R_{\ell}\leq R_{comb,\pi(\ell)}\triangleq \max\left(\frac{1}{2}\log\left(\frac{\mathrm{SNR}_{\ell}}{\| \mathbf{F}~\mathbf{a}_{\pi(\ell)}\|^2}\right),0\right) \label{c4}
\end{equation}
where the matrix $\mathbf{F}$ is given as
\vspace{-2mm}
\begin{equation*}
\mathbf{F}\triangleq\big(\frac{1}{P}\mathbf{I}_{K\times K}+\mathbf{h}\mathbf{h}^T\big)^{\frac{-1}{2}}\times\mathrm{diag}\left(\sqrt{\frac{\mathrm{SNR_1}}{P}},\dots,\sqrt{\frac{\mathrm{SNR_K}}{P}}\right).
\end{equation*}
The notation $\mathrm{diag}(\mathbf{v})$ stands for the diagonal matrix built from the vector $\mathbf{v}$ and $\mathrm{SNR}_{\ell}>0$ is the power used at encoder~$\ell$ to generate its codewords. Notice that as long as the generated codewords are scaled properly before transmission, they would satisfy the channel input power constraint.\footnote{The scaling factors can be absorbed into the the channel gains.}
\end{proposition}
Proposition~1 is immediately deduced from applying Theorem~2 along with Theorem~5 in~\cite{ordentlich2012approximate} with an exception that, here, users operate at different powers. All other conditions stated in Theorem~5 in~\cite{ordentlich2012approximate} still apply in Proposition~1.\\
\indent In the following, we present a lower bound on the secure sum capacity achieved by the proposed scheme.
\begin{theorem} \label{th1}
A rate tuple $\left(R_1,\dots,R_K\right)$ offers an achievable secure sum rate for the channel model described in Section~\ref{sec2}, if they satisfy the following constraints.
\vspace{-2mm}
\begin{equation}
 R_{\ell} \geq 0,~R_{\ell} \leq R_{comb,\pi(\ell)}  \quad \forall \ell \in \{1,\dots,K \} \label{13}
\end{equation}
\vspace{-2mm}
\begin{equation}\label{15b}
\sum _{\ell=1}^K R_{\ell} \leq \max_{\pi} R_{\textbf{sum}}~
\end{equation}
where
\vspace{-2mm}
\begin{equation}
R_{\textbf{sum}}=\left( \sum_{k=2}^K R_{comb,k}-\frac{1}{2}\log \bigg(\frac{\sum _{\ell=1}^K g_{\ell}^2}{g_{\pi^{-1}(1)}^2}\bigg) \right) \label{15}
\end{equation}
\end{theorem}
The maximum in~(\ref{15b}) is taken over all the possible successive cancellation orders~$\pi$ and the notation $\pi^{-1}(.)$ simply denotes the inverse permutation operator.\\
Proof of Theorem~1 is given in Section~\ref{sec6}.
\vspace{-1mm}
\subsection*{Comparison to the prior works}\label{sec4comp}
The $K$-user Gaussian wiretap MAC has been investigated in~\cite{tekin2008general} by means of i.i.d. Gaussian random coding. According to \cite{tekin2008general}, for the considered channel model, the following secure sum rate is achievable
\vspace{-2mm}
 \begin{equation}
\sum_{\ell=1}^K R_{\ell} \leq \max\left(\frac{1}{2}\log \bigg(\frac{1+\|\mathbf{h}\|^2P}{1+\|\mathbf{g}\|^2P}\bigg),0\right) \label{yenersum}
\end{equation}
\indent Note that the right hand side of expression~(\ref{yenersum}) does not scale with power $P$ or in other words, the asymptotic behavior of~(\ref{yenersum}) tends to a constant rate for a fixed number of users and a given set of channel gains. In contrast, our achievable secure sum rate in~(\ref{15}) scales logarithmic with $P$. To prove this, we only need to show that the first term in~(\ref{15}) grows with $\log(P)$ as the second term is constant with respect to the power. Without loss of generality, let us assume $\mathrm{SNR}_{\ell}=\alpha_{\ell}P, ~\forall~ \ell$ and some $\alpha_{\ell}>0$ (Note that according to the earlier discussion in Proposition~1,~$\alpha_{\ell}>1$ is allowed). Then we have,
\vspace{-5mm}
\begin{IEEEeqnarray}{l}
\nonumber \sum_{k=1}^K R_{comb,k}\\\nonumber
\stackrel{a}{\ge} \frac{K}{2}\log(P)+\frac{1}{2}\sum_{k=1}^K \log(\alpha_k)-\frac{1}{2}\log(K^K|\mathrm{det}(\mathbf{F})|^2)\\ \label{10p} \nonumber
=\frac{K}{2}\log(P)+\frac{1}{2}\sum_{k=1}^K \log(\alpha_k)-\frac{K}{2}\left(\log(K)+\log(P)\right)\\\nonumber
+\frac{1}{2}\log(1+\|\mathbf{h}\|^2P)-\frac{1}{2}\sum_{k=1}^K \log(\alpha_k)\\
=\frac{1}{2}\log(1+\|\mathbf{h}\|^2P)-\frac{K}{2}\log(K)\label{11p}
\end{IEEEeqnarray}
where inequality~(a) is deduced from Theorem~4 in~\cite{ordentlich2012approximate}. Now, we exploit Theorem~12 in~\cite{ordentlich2012approximate} in which it is shown that $R_{comb,k}<\frac{1+\delta(K-1)}{K+\delta(K-1)}.\frac{1}{2}\log(P)+c,~ \forall~k$, where the inequality holds for any $\delta>0$ and some $c$ constant with respect to $P$. Therefore, if we take $\delta\rightarrow 0$ and ignore the constant terms in~(\ref{11p}), we have $\sum_{k=2}^K R_{comb,k} \propto \frac{1}{2}.\frac{K-1}{K}\log(P)$. As a result, the secure sum rate in~(\ref{15}) grows with $\log(P)$.\\
\indent The numerical results are given in Fig.~2 which are evaluated for the three-user channel and random i.i.d. (real) Gaussian channel gains. It can be seen that for the moderate and high SNR regimes, our proposed scheme outperforms the random coding result presented in~\cite{tekin2008general}. Notice that the achievable non-secure results are shown in the figure as well which can be considered as an upper bound on the secure sum rate.\\
\indent Another interesting observation occurs when the channel to the legitimate receiver is degraded with respect to the channel to the eavesdropper. For the Gaussian setting and the same noise power, this corresponds to the case $\|\mathbf{h}\|\leq\|\mathbf{g}\|$. In this case, according to the expression in~(\ref{yenersum}), random coding fails to achieve a positive secure sum rate, while our scheme achieves a strictly positive secure sum rate as long as the ratios $\frac{\mathbf{h}_{\ell}}{\mathbf{g}_{\ell}}$ are not rational.\footnote{It can be shown that the Lebesgue measure of such rational ratios is small.} To illustrate this observation, we ran an experiment on a two-user Gaussian wiretap MAC with a fixed power (at SNR$=25\mathrm{dB}$) in which the channel gains are given as
\vspace{-4mm}
\begin{equation}
\mathbf{h}=
\left[1,\sqrt{2}\right]^T,~~\mathbf{g}=
\left[\sqrt{3}\cos(\theta),\sqrt{3}\sin(\theta)\right]^T \label{example}
\end{equation}
for some random $\theta$ uniformly distributed over $[0,2\pi]$. This is an example of the case where $\|\mathbf{h}\|=\|\mathbf{g}\|$. Fig.~3 shows that as long as the ratios of $\frac{h_\ell}{g_\ell}$ are not rational, a positive secure sum rate can be attained following our scheme.\\
\indent At last, we investigate the asymptotic behavior of the expression~(\ref{15}). We show that our scheme achieves a total secure DoF of $\frac{K-1}{K}$. Earlier, to prove the scalability of~(\ref{15}) with $\log(P)$, we showed that the $R_{\textbf{sum}}$ is proportional to $\frac{1}{2}.\frac{K-1}{K}\log(P)$, provided that the constant terms are ignored. Therefore,
\begin{equation}
 \lim_{P\rightarrow \infty} \frac{R_{\textbf{sum}}}{\frac{1}{2}\log \left(1+P\right)}=\frac{K-1}{K} \label{sdofp}
\end{equation} \label{dof}
Thus, the asymptotic behavior of the proposed scheme agrees with the result in~\cite{bagherikaram2010secure}. In fact, we can further improve the presented scheme so that its asymptotic behavior reaches the optimal secure degrees of freedom given in~\cite{xie2012secure}. The latter is aimed to be presented in the extended version.
\vspace{-0.3mm}
\begin{figure}
\vspace{-4mm}
\centering
\includegraphics[height= 1.45 in, width=0.5\textwidth]{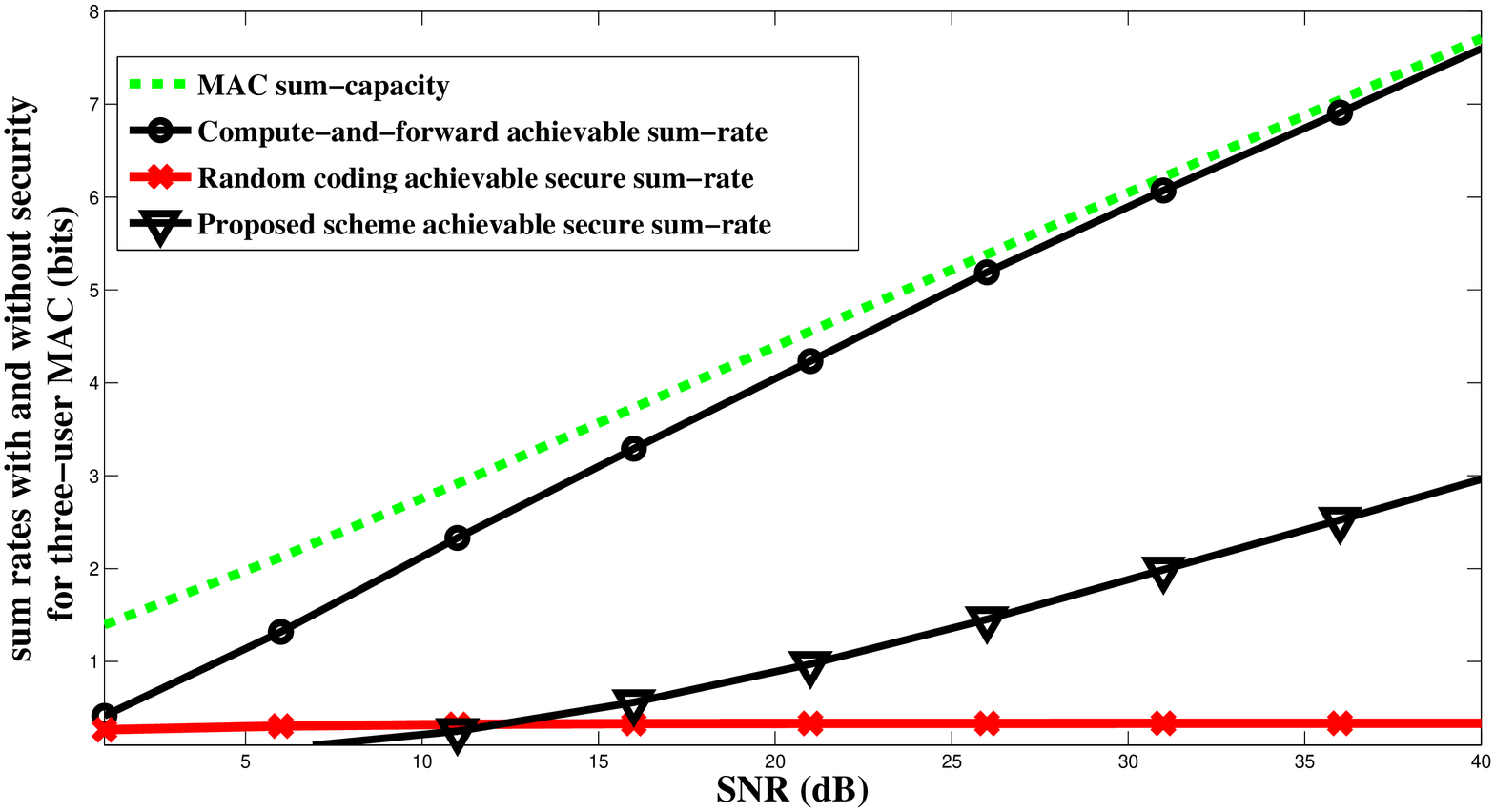}
\vspace{-7mm}
\centering
\caption{\small {Achievable sum rate, with and without security evaluated for the three-user asymmetric Gaussian MAC at different SNR.}}
\label{fig1}
\end{figure}
\begin{figure}[t]
\vspace{-4mm}
\centering
\includegraphics[height= 1.45 in, width=0.5\textwidth]{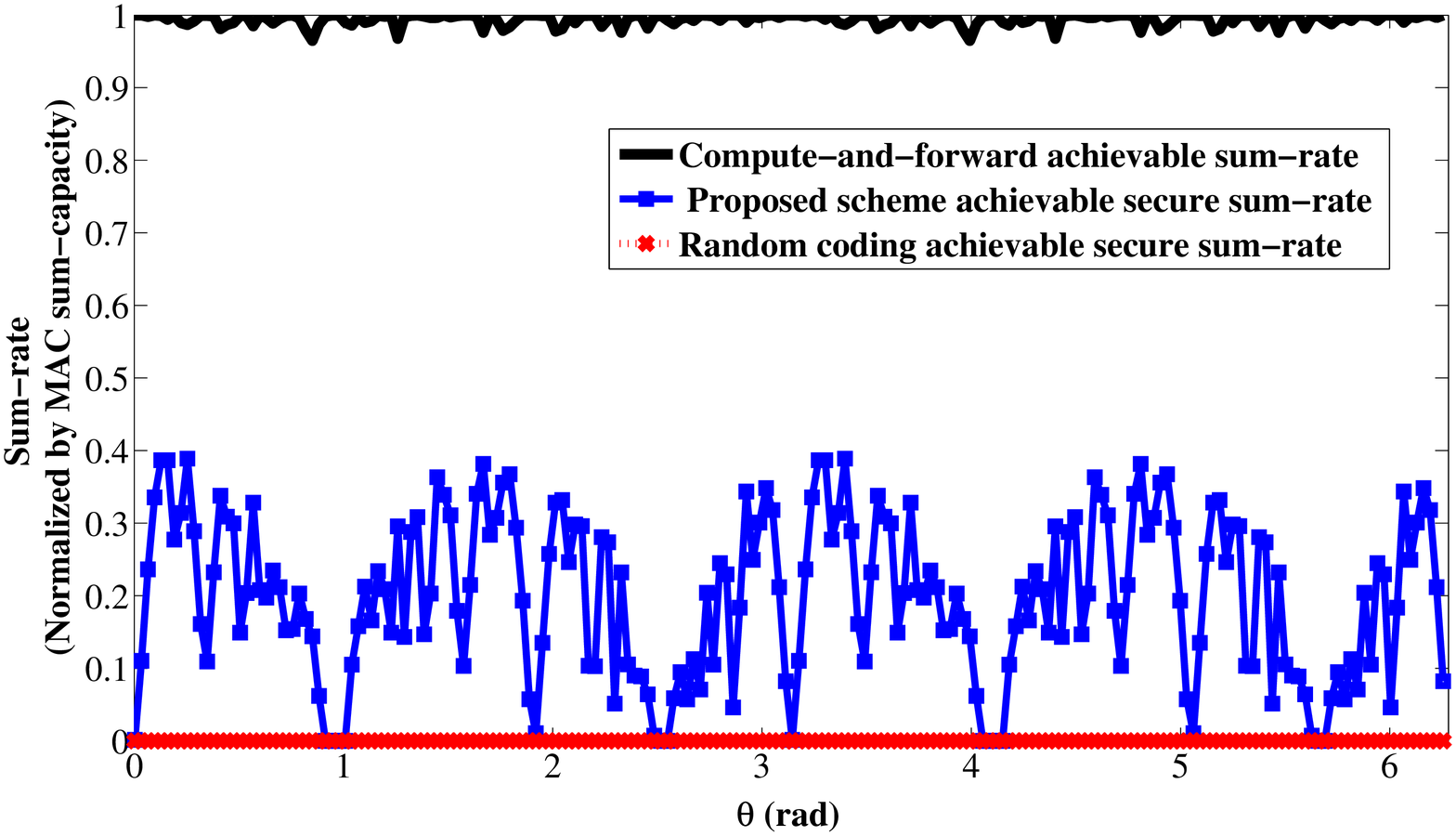}
\vspace{-5mm}
\centering
\caption{\small {Achievable sum rate evaluated for the two-user asymmetric Gaussian wiretap MAC with channel gains given as in~(\ref{example}) at SNR=25~dB.}}
\label{fig1}
\vspace{-5mm}
\end{figure}
\vspace{-1mm}
\section{Proof of Theorem~1}\label{sec6}
In this section, we use notions and properties related to the lattice coding and nested lattice structure which are discussed in detail in the seminal work by Erez and Zamir in~\cite{erez2004achieving}. Due to the space limitation, we avoid discussing the previously known results in this paper and we focus on the new results. Our proposed scheme provides security by confusing the eavesdropper through aligning the codewords at the eavesdropper side such that it can only decode the subsets of the codewords which have the same sum values in $\mathbb{R}^N$. To this end, each encoded codeword $\tilde{\mathbf{x}}$ at transmitter~$\ell$ is scaled before the transmission by a factor of $\frac{1}{g_{\ell}}$, i.e., $\mathbf{x}_{\ell}=\frac{\tilde{\mathbf{x}_{\ell}}}{g_{\ell}}$, so that the eavesdropper receives the sum of the codewords $\tilde{\mathbf{x}}$ as its channel output, i.e., $\mathbf{y}_E=\sm \tilde{\mathbf{x}}+\mathbf{z}_E$. Consequently, user~$\ell$ generates its codewords~$\tilde{\mathbf{x}}$ using power of $\mathrm{SNR}_{\ell}\triangleq g_{\ell}^2P$ so that the transmitted codewords $\mathbf{x}_{\ell}$ satisfy the power constraint in~(\ref{power}). \\
\indent As it was mentioned earlier, to address the problem of users with different powers, we utilize the asymmetric compute-and-forward framework along with a nested lattice structure. In our asymmetric compute-and-forward framework, user~$\ell$ generates a sequence of $n$-length lattice codewords ~$\mathbf{t}_{\ell}$ using a pair of fine and coarse lattice sets as $(\Lambda_{f,\ell},\Lambda_{\ell})$. The coarse lattice~$\Lambda_{\ell}$ is scaled such that its second moment equals to the available power at user~$\ell$, i.e., $\mathrm{SNR}_{\ell}=g_{\ell}^2P$. Also, we impose a nested structure on the users' lattice pairs as
\vspace{-1.8mm}
\begin{equation}
\Lambda_K \subseteq \Lambda_{K-1} \subseteq \dots \subseteq \Lambda_1 \subseteq \Lambda_{f,K} \subseteq \dots \subseteq \Lambda_{f,1} \label{nest}
\end{equation}
In the rest of the proof, we shall assume $\pi(\ell)=\ell,~\forall~\ell$ in~(\ref{15}). If that is not the case, we can simply re-index the users indices and define a nested structure as in~(\ref{nest}) for the re-indexed users.\\
\indent User~$\ell$ constructs its codebook in three steps. The first step for user~$\ell$ is to construct its inner codebook $\mathcal{L}_{\ell}\triangleq \Lambda_{f,\ell} \cap \mathcal{V}_{\ell}$, where $\mathcal{V}_{\ell}$ is the fundamental Voronoi region of the coarse lattice~$\Lambda_{\ell}$. The ratio between the coarse and the fine lattices is set such that~$\mathcal{L}_{\ell}$ consists of $2^{nR_{comb,\ell}}$ inner codewords $\mathbf{t}_{\ell}$, i.e., $R_{comb,\ell}=\frac{1}{n}\log \big|\Lambda_{f,\ell} \cap \mathcal{V}_{\ell}\big|,~\forall~\ell$. The inner codewords $\mathbf{t}_{\ell}$ have a uniform distribution over $\mathcal{L}_{\ell}$. \\
\indent In the second step, user~$\ell$ builds its outer codebook by generating $B$~i.i.d. copies of the inner codewords~$\mathbf{t}_{\ell}$, for some large enough $B$. Let us denote the outer codewords as $\bar{\mathbf{t}}_{\ell}$. Then we have $\bar{\mathbf{t}}_{\ell}\triangleq [\mathbf{t}_{\ell}^{[1]}, \dots, \mathbf{t}_{\ell}^{[B]}]$. Note that each $\mathbf{t}_{\ell}^{[i]}$ is independently and uniformly distributed over~$\mathcal{L}_{\ell}$. It is worth to mention that the outer code is added only for technicality reasons in the proof of Lemma~2 in~\cite{Extended} and it does not increase secrecy. Also, adding the outer layer to the codebook changes the block length of the overall codewords from $n$ to $N\triangleq B \times n$.\\
\indent Finally, in the third step, the wiretap codebook is built. To this end, user~$\ell$ partitions the outer codewords $\bar{\mathbf{t}}_{\ell}$ into $2^{NR_{\ell}}$ equal-size bins and randomly assigns each index $w_{\ell} \in \{1,\dots,2^{NR_{\ell}}\}$ to exactly one bin. Rates~$R_{\ell}$ are chosen such that they satisfy~(\ref{13}) and $\sm R_{\ell}=\sum_{\ell=2}^K R_{comb,\ell}-\frac{1}{2}\log \big(\frac{\sum _{\ell} g_{\ell}^2}{g_{1}^2}\big)+\epsilon_1$, for some small $\epsilon_1>0$. Also, user~$\ell$ has a random dither $\mathbf{d}^{[i]}_{\ell}$ for each block~$i$, which is independently generated according to a uniform distribution over $\mathcal{V}_{\ell}$. Dithers are public and do not increase secrecy.\footnote{As the average leakage rate (w.r.t. dithers) goes to zero, there must exist a sequence of deterministic dithers for which the leakage rate goes to zero.}\\
\indent To send a message $W_{\ell}=w_{\ell}$, user~$\ell$ randomly picks a codeword~$\bar{\mathbf{t}}_{\ell}$ from the corresponding bin and dithers it. Then, it scales the resulting codeword by the factor of $\frac{1}{g_{\ell}}$. The signal transmitted by user~$\ell$ is
\vspace{-2mm}
\begin{equation}
\mathbf{x}_{\ell}\stackrel{\bigtriangleup}{=}\frac{1}{g_{\ell}}\left(\left[\bar{\mathbf{t}}_{\ell}+\bar{\mathbf{d}}_{\ell}\right]\modnl\right) \label{send}
\end{equation}
Note that in~(\ref{send}) the modular operation is done block-wise, meaning that for $i \in \{1,\dots,B\}$ the signal transmitted at block~$i$ is $\frac{1}{g_{\ell}}([\mathbf{t}^{[i]}_{\ell}+\mathbf{d}^{[i]}_{\ell}]~\modnl)$.
\subsection*{Proof of secrecy}
\indent In this subsection, we bound the eavesdropper's equivocation rate. Without loss of generality, let us assume $R_{comb,\ell} >0,~\forall ~\ell$. We have
\vspace{-0.8mm}
\begin{IEEEeqnarray}{L}
\nonumber\frac{1}{N}H(W_1,\dots, W_K\big|\mathbf{y}_E,\bar{\mathbf{d}}_1,\dots,\bar{\mathbf{d}}_K)\\\nonumber
\geq\frac{1}{N}H(\bar{\mathbf{t}}_1,\dots,\bar{\mathbf{t}}_K \big|\mathbf{y}_E,\bar{\mathbf{d}}_1,\dots,\bar{\mathbf{d}}_K)\\\nonumber
- \frac{1}{N}H(\bar{\mathbf{t}}_1,\dots,\bar{\mathbf{t}}_K\big|W_1,\dots, W_K,\mathbf{y}_E,\bar{\mathbf{d}}_1,\dots,\bar{\mathbf{d}}_K)
\\\nonumber\stackrel{(a)}{\geq} \frac{1}{N}H(\bar{\mathbf{t}}_1,\dots,\bar{\mathbf{t}}_K \big|\mathbf{y}_E,\bar{\mathbf{d}}_1,\dots,\bar{\mathbf{d}}_K)-2\epsilon_2
\\\nonumber \geq\frac{1}{N}H(\bar{\mathbf{t}}_1,\dots,\bar{\mathbf{t}}_K \big|\mathbf{y}_E,\bar{\mathbf{d}}_1,\dots,\bar{\mathbf{d}}_K,\mathbf{z}_E)-2\epsilon_2
\\\nonumber \stackrel{(b)}{=}\frac{1}{N}H\left(\bar{\mathbf{t}}_1,\dots,\bar{\mathbf{t}}_K \bigg|\sm g_{\ell}\mathbf{x}_{\ell},\bar{\mathbf{d}}_1,\dots,\bar{\mathbf{d}}_K \right)-2\epsilon_2
\end{IEEEeqnarray}
\begin{IEEEeqnarray}{l}
\nonumber\stackrel{(c)}{=}\frac{1}{N}H\left(\bar{\mathbf{t}}_1,\dots,\bar{\mathbf{t}}_K \bigg|\left[\sm\bar{\mathbf{t}}_{\ell}\right]\modla, \bar{\mathbf{u}}_1,\bar{\mathbf{d}}_1,\dots,\bar{\mathbf{d}}_K \right)- 2\epsilon_2
\\\nonumber\stackrel{(d)}{=} \frac{1}{N}H\left(\bar{\mathbf{t}}_2,\dots,\bar{\mathbf{t}}_K \bigg|\left[\sm \bar{\mathbf{t}}_{\ell}\right]\modla,\bar{\mathbf{u}}_1,\bar{\mathbf{d}}_1,\dots,\bar{\mathbf{d}}_K \right)-2\epsilon_2
\\\nonumber \stackrel{(e)}{\geq} \frac{1}{N}H\left(\bar{\mathbf{t}}_2,\dots,\bar{\mathbf{t}}_K \bigg|\left[\sm \bar{\mathbf{t}}_{\ell}\right]\modla \right)-\frac{1}{N}H(\bar{\mathbf{u}}_1)-2\epsilon_2 \\\nonumber
\stackrel{(f)}{=} \frac{1}{N}H(\bar{\mathbf{t}}_2,\dots,\bar{\mathbf{t}}_K )-\frac{1}{N} H(\bar{\mathbf{u}}_1)-2\epsilon_2\\ \nonumber
\stackrel{(g)}{=}\frac{B}{N}\sum_{\ell=2}^KnR_{comb,\ell}- \frac{B}{N}H(\mathbf{u}^{[1]}_1) -2\epsilon_2\\\nonumber\label{dbound}
\stackrel{(h)}{\geq}\sum_{\ell=2}^KR_{comb,\ell}- (1-\epsilon)\frac{1}{2}\log\left(\frac{\sum_{\ell=1}^K g_{\ell}^2+ \epsilon}{g_1^2}\right)-\delta(\epsilon)-2\epsilon_2
\\\nonumber \geq \sum_{\ell=2}^KR_{comb,\ell}- \frac{1}{2}\log\left(\frac{\sum_{\ell=1}^K g_{\ell}^2}{g_1^2}\right)-\frac{\epsilon}{g_1^2}-\delta(\epsilon)-2\epsilon_2\\
\stackrel{(i)}{=} \sum_{\ell=2}^KR_{comb,\ell}- \frac{1}{2}\log\left(\frac{\sum_{\ell=1}^K g_{\ell}^2}{g_1^2}\right)-\epsilon_3
\end{IEEEeqnarray}
\vspace{-1mm}
In the above inequalities, (a) is deduced from applying the packing lemma to the outer codewords (detailed proof of this
step is provided in Appendix of \cite{Extended}). (b) is true since after subtracting the noise from $\mathbf{y}_E$, the remaining random vectors become independent of the noise. (c) is true since $\Lambda_1$ is the densest lattice among the lattices~$\left(\Lambda_1,\Lambda_2,\dots,\Lambda_K\right)$, according to the nested structure in~(\ref{nest}). Therefore,
\begin{equation*}
\left[\sm g_{\ell}\mathbf{x}_{\ell}-\sm \bar{\mathbf{d}}_{\ell}\right]\modla
= \left[\sm\bar{\mathbf{t}}_{\ell}\right] \modla.
\end{equation*}
Also, notice that
\begin{equation*}
H\left(\sm g_{\ell}\mathbf{x}_{\ell}\right)=H\left(\left[\sm g_{\ell}\mathbf{x}_{\ell}\right]\modla,\bar{\mathbf{u}}_1\right),
\end{equation*}
where $\bar{\mathbf{u}}_1 \stackrel{\bigtriangleup}{=} \sm g_{\ell}\mathbf{x}_{\ell} - \left[\sm g_{\ell}\mathbf{x}_{\ell} \right] \modla$.
Inequality (d) is due to the reason that the codeword $\bar{\mathbf{t}}_1$ can be obtained from the modulo-sum $ \left[\sm \bar{\mathbf{t}}_{\ell}\right]\modla$ and the sequence of codewords $\bar{\mathbf{t}}_{2}, \dots, \bar{\mathbf{t}}_{K}$. (e) holds since dithers are independent of the codewords and conditioning reduces entropy. (f) is deduced from Lemma 2 in~\cite{forney2003role} (Crypto lemma), which implies that for each block ~$i \in [1,B]$, $\left[\mathbf{t}_1^{[i]}+\sum_{\ell=2}^K \mathbf{t}_{\ell}^{[i]}\right]\modla$ has uniform distribution over the codebook $\mathcal{L}_1$ and is independent of $\sum_{\ell=2}^K \mathbf{t}_{\ell}^{[i]}$. (g) is true since $\big|\mathcal{L}_{\ell}\big|=2^{nR_{comb,\ell}}$ and for $i \in\{1,\dots,B\}$, inner codewords $\mathbf{t}^{[i]}_{\ell}$ have i.i.d. uniform distribution over $\mathcal{L}_{\ell},~\forall~\ell$. Also, $\bar{\mathbf{u}}_1$ consists of $B$~i.i.d. copies of $\mathbf{u}^{[i]}_1$ by its definition. (h) follows from applying Lemma 1 in Appendix to $\mathbf{u}^{[1]}_1$, and finally, (i) is deduced by defining $\epsilon_3 \triangleq \delta(\epsilon)+\frac{\epsilon}{g_1^2}+2\epsilon_2$. Thus, the condition in~(\ref{c3}) is satisfied and the proof of secrecy is completed.
\vspace{-1mm}
\section{Conclusion}\label{sec7}
In this paper, we proposed a security scheme built on the asymmetric compute-and-forward framework, which works at any finite SNR. The achievable secure sum rate presented in our scheme scales with $\log(\mathrm{SNR})$ and therefore, it significantly outperforms the existing random coding result for the most SNR regimes. Our presented scheme also achieves a total secure DoF of~$\frac{K-1}{K}$. This result can be furthered improved to achieve the optimal secure DoF which is aimed to be presented in our future work.
\vspace{-3mm}
\section*{Acknowledgment}
The authors would like to thank Bobak Nazer and Prakash Ishwar for their valuable comments and helpful discussions.
\bibliography{refn}
\vspace{-2mm}
\section*{Appendix}
\begin{lemma} Consider a set of $n$-dimensional lattices $\Lambda_1,\dots,\Lambda_K$ with their fundamental Voronoi regions as $\mathcal{V}_1,\dots,\mathcal{V}_K$, respectively. Assume that all the lattices are scaled such that their second moments equal to $\mathrm{SNR}_\ell=g_{\ell}^2P, ~\forall ~\ell \in \{1,\dots,K\}$, where $P>0$. Now construct random vectors $\mathbf{u}_j$, for $j\in \{1,\dots , K\}$, as $\mathbf{u}_j\triangleq Q_{\Lambda_j}\left(\sm \mathbf{s}_{\ell}\right)$, where $\mathbf{s}_1,\dots,\mathbf{s}_K$ are independent $n$-dimensional random vectors uniformly distributed over $\mathcal{V}_1,\dots,\mathcal{V}_K$, respectively, and the operation $Q_{\Lambda_j}(.)$ is the nearest neighbor quantizer with respect to the lattice~$\Lambda_j$. Then, for all $ \epsilon >0$ and sufficiently large~$n$, the entropy of $\mathbf{u}_j$ is bounded as
\vspace{-2mm}
\begin{equation}
\frac{1}{n}H(\mathbf{u}_j)\leq (1-\epsilon)\frac{1}{2}\log\left(\frac{\sum_{\ell=1}^K g_{\ell}^2+ \epsilon}{g_j^2}\right)+\delta(\epsilon) \quad \forall j 
\end{equation}
where $\delta(\epsilon)$ tends to zero as $\epsilon\rightarrow 0$.
\end{lemma}
~~~~\textit{Proof:} According to Lemma~1, $\mathbf{u}_j$ is the output of the lattice quantizer~$Q_{\Lambda_j}$, so it can only take discrete values. To bound the entropy of $\mathbf{u}_j$, first we bound the range of $\|\sm \mathbf{s}_{\ell}\|$ as follows. Let $r_{\mathrm{cov},\ell}$ denote the covering radius of $\Lambda_{\ell}$, i.e., the radius of the smallest ball containing the Voronoi region~$\mathcal{V}_{\ell}$. Also, let $r_{\mathrm{eff},\ell}$ denote the radius of the sphere which has the same volume as the volume of $\mathcal{V}_{\ell}$, i.e., $\mathrm{Vol}(B(r_{\mathrm{eff},\ell}))=\mathrm{Vol}(\mathcal{V}_{\ell})$. Now, consider $K$ ($n$-dimensional) balls whose second moments per dimension equal to $\sigma_1^2,\sigma_2^2,\dots,\sigma_K^2$ and their radii are given as $r_{\mathrm{cov},1},r_{\mathrm{cov},2},\dots,r_{\mathrm{cov},K}$, respectively. Next, for each $\ell \in \{1,\dots ,K\}$, consider a random vector $\mathbf{b}_{\ell}$ with the uniform distribution over an $n$-dimensional ball $B(r_{\mathrm{cov},\ell})$. Recall that a ball has the smallest normalized second moment for a given volume~\cite{erez2004achieving}. Therefore, we have
\vspace{-2mm}
\begin{IEEEeqnarray}{l}
\nonumber g_{\ell}^2P=\frac{1}{n}\mathbb{E}\|\mathbf{s}_{\ell}\|^2
\\ \geq \frac{1}{n}\mathbb{E}\bigg\|\rf \mathbf{b}_{\ell}\bigg\|^2=\rf^2\sigma_{\ell}^2,~\forall~\ell \label{rr}
\end{IEEEeqnarray}
Now, consider a random vector $\mathbf{z}_{eq}\triangleq \sm \mathbf{z}_{\ell}$, in which random vectors $\mathbf{z}_{\ell}$ are i.i.d. according to the distribution~$\mathcal{N}(\mathbf{0},\sigma_{\ell}^2\mathbf{I})$ and therefore, $\mathbf{z}_{eq}\sim \mathcal{N}(\mathbf{0},\sigma_{eq}^2\mathbf{I})$. Then, from~(\ref{rr}) we have $\sigma_{\mathbf{z}_{eq}}^2=\sum_{\ell=1}^K\sigma_{\ell}^2\leq \sum_{\ell=1}^K \left(\frac{r_{\mathrm{cov},\ell}}{r_{\mathrm{eff},\ell}}\right)^2 g_{\ell}^2P$.
Now, using Lemma 11 in~\cite{erez2004achieving}, we conclude that
\begin{IEEEeqnarray}{l}
\nonumber e^{K.n.c(n)}f_{\mathbf{z}_{eq}}(\mathbf{z}_{eq})=e^{K.n.c(n)}\left(f_{\mathbf{z}_1}(\mathbf{z}_{eq})*...*f_{\mathbf{z}_K}(\mathbf{z}_{eq})\right)\\
\geq f_{\sum_{\ell=1}^K \mathbf{s}_{\ell}}(\mathbf{z}_{eq}) \label{zq}
\end{IEEEeqnarray}
where $n.c(n)$ goes to zero as $n$ goes to infinity. Notice that in deriving~(\ref{zq}) we also used the fact that vectors $\mathbf{s}_{\ell}$ are independent vectors, and hence, pdf of their sum is the convolution of their individual pdfs. Now we can bound the range of $ \|\sum_{\ell=1}^K \mathbf{s}_{\ell}\|$ as follows. For any $\epsilon>0$,
\vspace{-1.5mm}
\begin{IEEEeqnarray}{l}
\nonumber \mathrm{Pr}\bigg(\big \|\sum_{\ell=1}^K \mathbf{s}_{\ell}\big\| \not \in B\left(\sqrt{n\sigma^2_{\mathbf{z}_{eq}}+n\epsilon}\right)\bigg)\\\nonumber
\stackrel{(a)}{\leq} e^{K.n.c(n)} \mathrm{Pr}\bigg(\|\mathbf{z}_{eq}\| \not \in B\left(\sqrt{n\sigma^2_{\mathbf{z}_{eq}}+n\epsilon}\right)\bigg)
\leq \epsilon. 
\end{IEEEeqnarray}
Inequality~(a) follows from~(\ref{zq}) and non-negativity of the $\ell_2$-norm. Also, the last inequality is deduced from the Weak Law of Large numbers (WLL) for sufficiently large~$n$. Since we showed that $\big \|\sum_{\ell=1}^K \mathbf{s}_{\ell}\big\|$ belongs to the ball $B(\sqrt{n\sigma_{zeq}^2+n\epsilon})$ with probability $1-\epsilon$, it only remains to find an upper bound on the number of non-intersecting Voronoi regions $\mathcal{V}_j$ which fit in this ball, i.e.,
\begin{IEEEeqnarray*}{l}
\nonumber \frac{Vol \bigg( \mathcal{B}\left(\sqrt{n\sigma_{\mathbf{z}_{eq}}^2+n\epsilon}\right) \bigg)}{Vol\left(\mathcal{V}_j\right)}=\frac{Vol \bigg( \mathcal{B}\left(\sqrt{n\sigma_{\mathbf{z}_{eq}}^2+n\epsilon}\right) \bigg)}{Vol\left(B(r_{\mathrm{eff},j})\right)} \\
\stackrel{(a)}{\leq} \bigg(\frac{n\sigma_{\mathbf{z}_{eq}}^2+n\epsilon}{ \left(\frac{r_{\mathrm{eff},j}}{r_{\mathrm{cov},j}}\right)^2 ng_j^2P}\bigg)^{\frac{n}{2}}
\stackrel{(b)}{\leq} \bigg(\frac{\sm \left(\frac{r_{\mathrm{cov},\ell}}{r_{\mathrm{eff},\ell}}\right)^2 g_{\ell}^2+\epsilon}{\left(\frac{r_{\mathrm{eff},j}}{r_{\mathrm{cov},j}}\right)^2g_j^2}\bigg)^{\frac{n}{2}},
\end{IEEEeqnarray*}
where inequality (a) is concluded from Lemma~6 in~\cite{erez2004achieving} and inequality 
(b) follows from~(\ref{rr}). Finally, recall that for a high dimensional good lattices, we have $\log\left(\frac{r_{\mathrm{cov},\ell}}{ r_{\mathrm{eff},\ell}}\right) \rightarrow 0$~\cite{erez2004achieving}. Therefore,
\vspace{-2mm}
\begin{IEEEeqnarray}{l}
\nonumber \frac{1}{n}H(\mathbf{u}_j) \leq(1-\epsilon)\frac{1}{2}\log \bigg(\frac{\sm g_{\ell}^2+\epsilon}{g_j^2}\bigg)+\delta(\epsilon).\label{axiom}
\vspace{-2mm}
\end{IEEEeqnarray}
Note that using WLL, the term $\delta(\epsilon)$ tends zero as $n$ goes to infinity. This completes the proof.\hfill $\blacksquare$

\end{document}